\pdfoutput=1
\documentclass[5p,number,sort&compress,times,11pt]{elsarticle}
\usepackage{graphicx,epsfig,float}
\usepackage{dcolumn}
\usepackage{bm}    
\usepackage{amssymb} 
\usepackage{amsmath}
\usepackage{verbatim}
\usepackage{footnote}

\begin{document}
\begin{frontmatter}
\title{Edge stacking dislocations in two-dimensional bilayers with a small lattice mismatch}

\author[b]{Irina V. Lebedeva\corref{cor}}
\ead{liv\_ira@hotmail.com}
\address[b]{Nano-Bio Spectroscopy Group and ETSF, Universidad del Pa\'is Vasco, CFM CSIC-UPV/EHU, 20018 San Sebastian, Spain
}
\cortext[cor]{Corresponding author}
\author[a]{Andrey A. Knizhnik}
\ead{knizhnik@kintechlab.com}
\address[a]{Kintech Lab Ltd., 3rd Khoroshevskaya Street 12, Moscow 123298, Russia}
\author[c]{Andrey M. Popov}
\ead{popov-isan@mail.ru}
\address[c]{Institute for Spectroscopy of Russian Academy of Sciences, Troitsk, Moscow 108840, Russia}

\begin{abstract}
Incomplete stacking dislocations are predicted to form at edges of the shorter upper layer in two-dimensional hexagonal bilayers upon stretching the longer bottom layer. A concept of the edge Burgers vector is introduced to describe such dislocations by analogy with the Burgers vector of standard bulk dislocations. Analytical solutions for the structure and energy of edge stacking dislocations in bilayer graphene are obtained depending on the magnitude of elongation and angles between the edge Burgers vector, direction of elongation and edge. The barrier for penetration of stacking dislocations inside the bilayer is estimated. The possibilities to measure the barrier to relative motion of graphene layers and strain of graphene on a substrate by observation of edge stacking dislocations are discussed.      
\end{abstract}
\end{frontmatter}

\section{Introduction}
\begin{figure}
	\centering
	\includegraphics[width=\columnwidth]{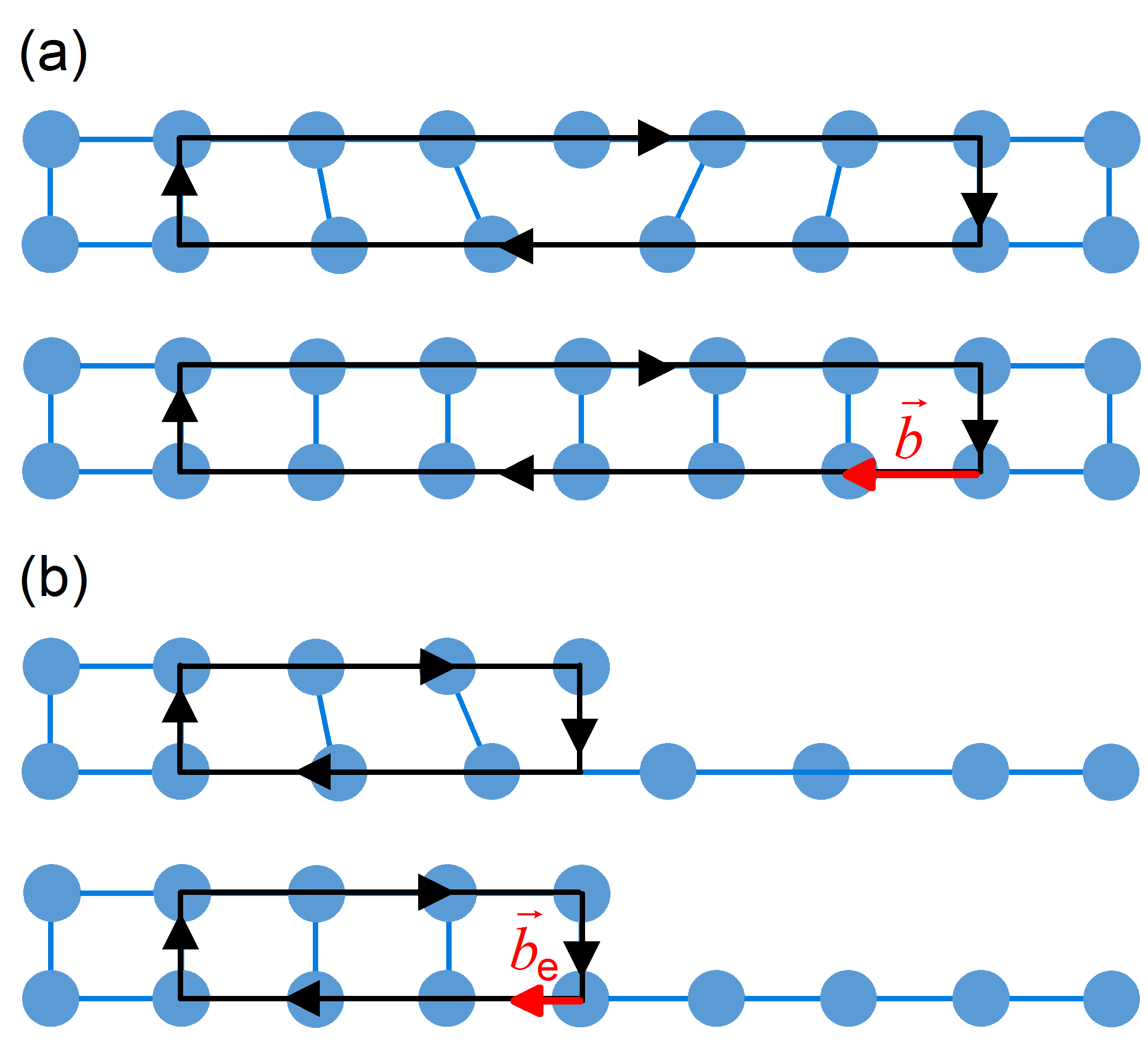}
	\caption{Schematic representation of stacking dislocations inside (a) and at the edge (b) of a bilayer. The standard bulk (a) and edge (b) Burgers vectors $\vec{b}$ and $\vec{b}_\mathrm{e}$ are indicated.}
	\label{fig:scheme}
\end{figure}

Dislocations associated with a variation in stacking of two-dimensional hexagonal layers (Fig.~\ref{fig:scheme}a) and manifested through incommensurate boundaries between commensurate domains have been recently in focus of extensive experimental \cite{Alden2013,Butz2014,Lin2013,Yankowitz2014} and theoretical \cite{Lin2013,Popov2011,Lebedev2015,Lebedeva2016} research. Similar to in-plane defects, such stacking dislocations affect electronic \cite{Hattendorf2013, San-Jose2014, Lalmi2014, Benameur2015, Koshino2013} and optical \cite{Gong2013} properties of bilayer and few-layer systems and, therefore, have implications for development of nanoelectronic devices. 

In addition to stacking dislocations occasionally present in bilayer and few-layer samples, it has been proposed that they can be intentionally generated by stretching of one of the layers. While at small external strains the interlayer interaction keeps the layers commensurate, increasing the strain to some critical value leads to a release of the excessive elastic energy through formation of dislocations. It has been shown that this is 
the second-order phase transition characterized by the density of stacking dislocations as the order parameter \cite{Pokrovsky1978,Popov2011}. Predicted theoretically for hexagonal bilayers \cite{Popov2011,Lebedeva2016} and nanotubes with commensurate walls \cite{Bichoutskaia2006,Popov2009}, this new phenomenon, commensurate-incommensurate phase transition, is still waiting for the experimental validation. In the meanwhile, the crossover from the structure with commensurate domains separated by incommensurate boundaries to the fully incommensurate state has already been observed for the layers with a small mismatch of the lattice constants upon changing the relative orientation of the layers \cite{Woods2014}.

Furthermore, physical phenomena where stacking between two-dimensional hexagonal layers is different from the ground-state one and is related to the interaction between the edges  have been observed recently. Bilayer graphene with the AA stacking and common folded edge where adjacent layers form a curved closed loop has been fabricated by heat treatment \cite{liu2009}. Lock-in positions of a graphene flake on the underlying graphene layer with the stacking different from the AB stacking and resulting from the edge-edge interaction have been studied by high-resolution electron microscopy \cite{cruz-silva2013}.

\begin{figure}
	\centering
	\includegraphics[width=\columnwidth]{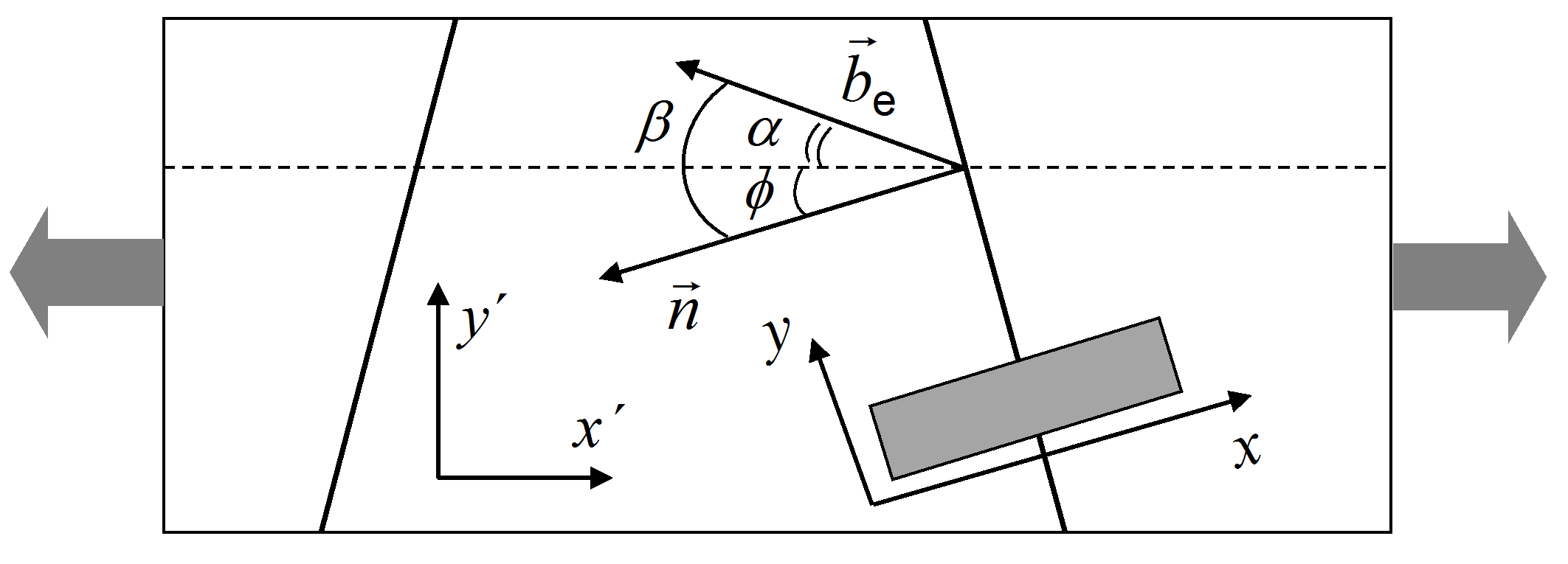}
	\caption{Scheme of formation of stacking dislocations at edges (inclined lines) of the upper layer of bilayer graphene upon elongation of the bottom layer (rectangular). The normal $\vec{n}$ to the edge and edge Burgers vector $\vec{b}_\mathrm{e}$ are shown. The direction of elongation is indicated by the thick arrows and dashed line. $\alpha$ is the angle between the direction of elongation and edge Burgers vector $\vec{b}_\mathrm{e}$. $\phi$ is the angle between the normal $\vec{n}$ to the edge and direction of elongation. $\beta = \alpha + \phi$ is the angle between the normal $\vec{n}$ to the edge and edge Burgers vector $\vec{b}_\mathrm{e}$. The coordinate systems $x-y$ and $x'-y'$ associated with the edge and elongation applied, respectively, are shown.} 
	\label{fig:statement_edge}
\end{figure}

In the present paper we propose yet another edge phenomenon which could be observed in two-dimensional hexagonal bilayers with a strained layer. Different from the previous studies dealing with stacking dislocations inside bilayers, we consider evolution of stacking at the edge of the shorter layer on top of the longer layer that is being stretched (Fig.~\ref{fig:statement_edge}). We suggest that any non-zero elongation of the bottom layer results in formation of an incomplete stacking dislocation at the edge of the upper layer (Fig.~\ref{fig:scheme}b), hereafter referred to as an edge stacking dislocation (ESD). The consideration of the edge evolution also allows us to estimate the barriers for penetration of complete stacking dislocations inside the layers and, therefore, to get an insight into kinetics of dislocation generation. Similar to transmission electron microscopy studies of pre-existing dislocations in few-layer graphene that resulted in an experimental estimate of the barrier to relative sliding of graphene layers \cite{Alden2013}, experimental investigation of ESDs could help to validate the theoretical predictions and learn more about interaction of atomically thin layers. Particularly we suggest possible experimental schemes using edge stacking dislocations for measurements of the barrier to relative motion of graphene layers, strain of graphene on a subtrate and average lattice constants of graphene membranes chemically modified at one side.

\begin{figure}
	\centering
	\includegraphics[width=\columnwidth]{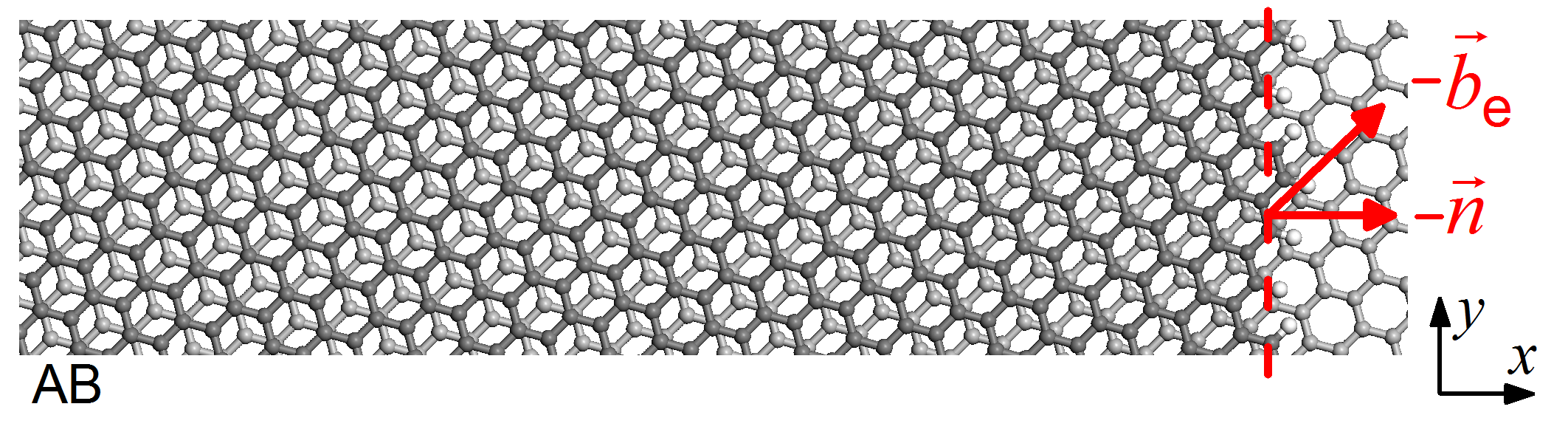}
	\caption{Atomistic structure corresponding to the stacking dislocation at the edge of the upper layer of bilayer graphene with the angle  $\beta = 45^{\circ}$ between the edge Burgers vector $\vec{b}_\mathrm{e}$ and normal $\vec{n}$ to the edge (Fig.~\ref{fig:statement_edge}) for the elongation of the bottom layer of $\epsilon = 6.28 \cdot 10^{-3}$ perpendicular to the edge ($\phi=0^{\circ}$ and $\alpha=\beta$). Carbon atoms of the upper and bottom layers and hydrogen atoms are coloured in dark gray, light gray and white, respectively. The magnitude of the vector $\vec{b}_\mathrm{e}$ is scaled up for clarity. The coordinate system $x-y$ associated with the edge is indicated.}
	\label{fig:struct}
\end{figure}

Special attention has been paid so far to dislocations in graphene \cite{Popov2011,Alden2013,Butz2014,Lin2013,Yankowitz2014,Hattendorf2013, San-Jose2014, Lalmi2014, Benameur2015, Koshino2013,Gong2013}. The potential surface of interlayer interaction energy in bilayer graphene has two degenerate inequivalent minima AB and AC. The transition between these two minima corresponds to formation of partial stacking dislocations (PSDs) with the Burgers vector $\vec{b}$ equal in magnitude to the bond length $l$ and smaller than the lattice constant $a_0 = l \sqrt{3}$. In addition to graphene, predictions regarding the properties of dislocations have been also made for hexagonal boron nitride. In particular, it has been suggested that PSDs similar to the ones in graphene could be also found in metastable boron nitride with co-aligned layers (AB stacking in the commensurate state) \cite{Lebedev2015,Lebedeva2016}, which is only slightly unstable compared to the ground state with the layers aligned in the opposite directions \cite{Lebedev2015} and has been observed experimentally \cite{Warner2010}. In the present paper, we consider graphene as an example (Fig.~\ref{fig:struct}). However, our conclusions are qualitatively valid for other hexagonal bilayers, while the results for properties of dislocations in boron nitride with the layers aligned in the same direction and graphene are close even quantitatively \cite{Lebedev2015}.

\section{Results}
We consider the case when the upper adsorbed layer is somewhat smaller than the bottom layer to avoid the interaction between their edges. The bottom layer is stretched, while the upper layer is left to relax freely. We assume that there is no difference in the interlayer interaction energy inside the layers and at the edge of the upper layer, which is the case, for example, when edges of the upper layer are terminated by hydrogen (Fig.~\ref{fig:struct}). This ensures that the formalism of the two-chain Frenkel-Kontorova model \cite{Bichoutskaia2006} used previously to study dislocations inside double-walled carbon nanotubes \cite{Bichoutskaia2006, Popov2009}, graphene \cite{Popov2011,Lebedeva2016} and boron nitride \cite{Lebedev2015,Lebedeva2016} can still be applied.  In this model, the upper and bottom layers are represented by two chains of particles connected by harmonic springs and coupled through van der Waals interactions. To consider ESDs we suppose that particle-spring pairs correspond to ribbons of the layers parallel to the edge. The edges are assumed to be straight so that relative displacements of atoms in the layers at each edge of the upper layer depend only of the coordinate $x$ along the normal to the edge. We also restrict our consideration to the case when the density of dislocations is low so that the interaction between ESDs and PSDs inside the layers can be neglected. 

Let us first review the main conclusions of the two-chain Frenkel-Kontorova model for PSDs in graphene \cite{Popov2011,Lebedev2015,Lebedeva2016}.
The first conclusion is that the path of PSDs, i.e. the dependence of relative displacement of the layers $ \vec{u}(x)$ on the coordinate $x$ perpendicular to the boundary between commensurate domains that minimizes the formation energy, lies along the straight line between adjacent energy minima AB and AC. Therefore, in this case the dislocation path is parallel to the Burgers vector, which describes the change of the relative displacement of the layers at the final and initial points of the dislocation path (Fig.~\ref{fig:scheme}a), at any point along the path. In the limit of an isolated PSD, the magnitude of the relative displacement of the layers $u(x)$ along the dislocation path (or along the direction of the Burgers vector) is determined by equality in the densities of the elastic and interlayer interaction energies
\begin{equation} \label{eq_1}
\begin{split}
\frac{1}{4} K(\beta) l^2 |u^{\prime}|^2 = V(u), 
\end{split}
\end{equation}
where $u$ is in units of the bond length $l$ and changes from 0 to $u_\mathrm{e} = 1$, which corresponds to the change in the relative displacements of the layers over the boundary between commensurate domains, $V(u)$ is the interlayer interaction energy per unit area, $|u^{\prime}| =\mathrm{d} u/\mathrm{d}x$ is the relative strain in the layers associated with formation of the dislocation, $\beta$ is the angle between the Burgers vector and normal to the boundary between commensurate domains, $K(\beta) = E \cos^2{\beta} + G \sin^2 {\beta}$ describes the dependence of the elastic constant on fractions of tensile and shear character in the dislocation, $E = k/(1-\nu^2)$ and $G = k/2(1+\nu)$ are the tensile and shear elastic constants per unit area, respectively,  $k = Yd$ is the elastic constant under uniaxial stress and $\nu$ is the Poisson ratio. In the present paper, we use  the following parameters for graphene obtained by density functional theory calculations \cite{Lebedeva2016} using the vdW-DF2 functional \cite{Lee2010}: $l = 1.430$~\AA, $k = 331 \pm 1$~J/m$^2$, $\nu = 0.174 \pm 0.002$ and the barrier to relative sliding of graphene layers $V_\mathrm{max} = 1.61$~meV/atom (in meV per atom in the upper/adsorbed layer).

It can be noted that consideration of a pair of chains only one of which is infinite and the other one is semi-infinite in the Frenkel-Kontorova model instead of two infinite chains  does not change the conditions on the dislocation path. For bilayer graphene, the dislocation path is still straight and satisfies Eq.~(\ref{eq_1}) in the presence of the edge. However, in such cases, solutions with the limits $u=0$ and $u_\mathrm{e} < 1$ inside the commensurate domain and at the edge become possible and can be attributed to ESDs. The condition $u_\mathrm{e} < 1$ means that an ESD is incomplete in the sence that only a part of the path between the minima on the potential energy surface is made. 

\begin{figure}
	\centering
	\includegraphics[width=\columnwidth]{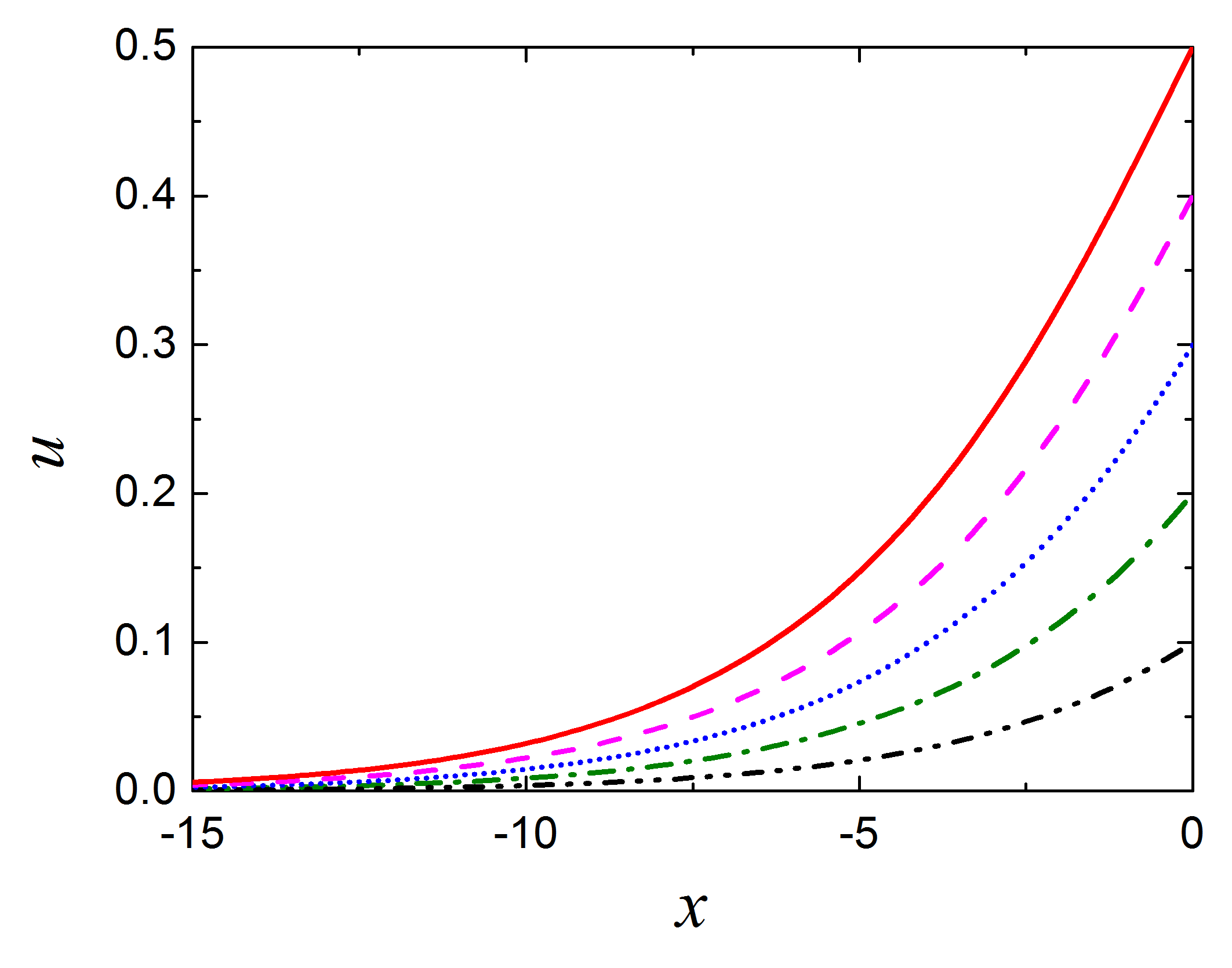}
	\caption{Relative displacement $u$  (in units of the bond length $l$) of atoms of the upper and bottom layers along the dislocation path as a function of their position $x$ (in nm) in the direction of the normal $\vec{n}$ to the edge for the edge stacking dislocation in bilayer graphene with the angle $\beta = 45^{\circ}$ between the edge Burgers vector $\vec{b}_\mathrm{e}$ and normal $\vec{n}$ to the edge for the bottom layer (Fig.~\ref{fig:statement_edge}) stretched perpendicular to the edge ($\phi=0^{\circ}$ and $\alpha=\beta$). Elongations $\epsilon$ of the bottom layer are  $6.57 \cdot 10^{-3}$ (red solid line), $6.28 \cdot 10^{-3}$ (magenta dashed line), $5.43 \cdot 10^{-3}$ (blue dotted line),  $4.06 \cdot 10^{-3}$ (green dash-dotted line) and  $2.22 \cdot 10^{-3}$ (black dash-dot-dotted line). The coordinate $x=0$ corresponds to the position of the edge. In the coordinate system associated with the edge the relative displacements of atoms of the layers are $u_x=-u\cos{\beta}$ and $u_y=-u\sin{\beta}$.}
	\label{fig:path}
\end{figure}

For ESDs, we assume that the relative displacement of the layers is the function of the coordinate $x$ perpendicular to the edge. The ``edge" Burgers vector can be introduced by analogy with the standard bulk Burgers vector for PSDs so that it describes the change of the relative displacement of the layers at the edge and inside the commensurate domain (Fig.~\ref{fig:scheme}b) and changes continuously from 0 to $l$. Therefore, in the case of ESDs, $\beta$ is the angle between the edge Burgers vector and normal to the edge (Fig.~\ref{fig:statement_edge}). Examples of dislocation paths of ESDs in bilayer graphene are shown in Fig.~\ref{fig:path}.

Based on Eq.~(\ref{eq_1}) the formation energy of PSDs, i.e. the energy of the state with a dislocation relative to the fully commensurate state, is represented through the sum of two terms \cite{Popov2011,Lebedev2015,Lebedeva2016}. The first term describes the formation energy in the absence of external strain and can be expressed via the geometric mean of  the densities of the elastic and interlayer interaction energies, while the second term corresponds to the energy gain due to the elongation of the bottom layer. The formation energy per unit length of the boundary between commensurate domains then takes the form
\begin{equation} \label{eq_2}
\begin{split}
\Delta W = & \int\limits_{0}^{u_\mathrm{e}} \sqrt{ K(\beta) l^2 V(u)}\mathrm{d}u  -\epsilon kl u_\mathrm{e} \cos{\alpha} \cos{\phi},
\end{split}
\end{equation} 
where $\epsilon$ is the relative elongation of the bottom layer, $\alpha$ is the angle between the direction of elongation and Burgers vector, $\phi = \beta - \alpha$ is the angle between the direction of elongation and normal to the boundary between commensurate domains and $u_\mathrm{e} = 1$ for PSDs, as mentioned before. 

This equation is still valid for ESDs with the only difference that the relative displacement $u_\mathrm{e}$ of the layers at the edge of the upper layer is smaller in magnitude than unity, $u_\mathrm{e} < 1$. The angles $\alpha$ and $\phi = \beta - \alpha$ for this case are shown in Fig.~\ref{fig:statement_edge}. It should be noted that from Eq.~(\ref{eq_2}) it is already clear that the elongation of the bottom layer has no effect on the formation energy $\Delta W$ when the direction of elongation is parallel to the edge or perpendicular to the edge Burgers vector.

\begin{figure}
	\centering
	\includegraphics[width=\columnwidth]{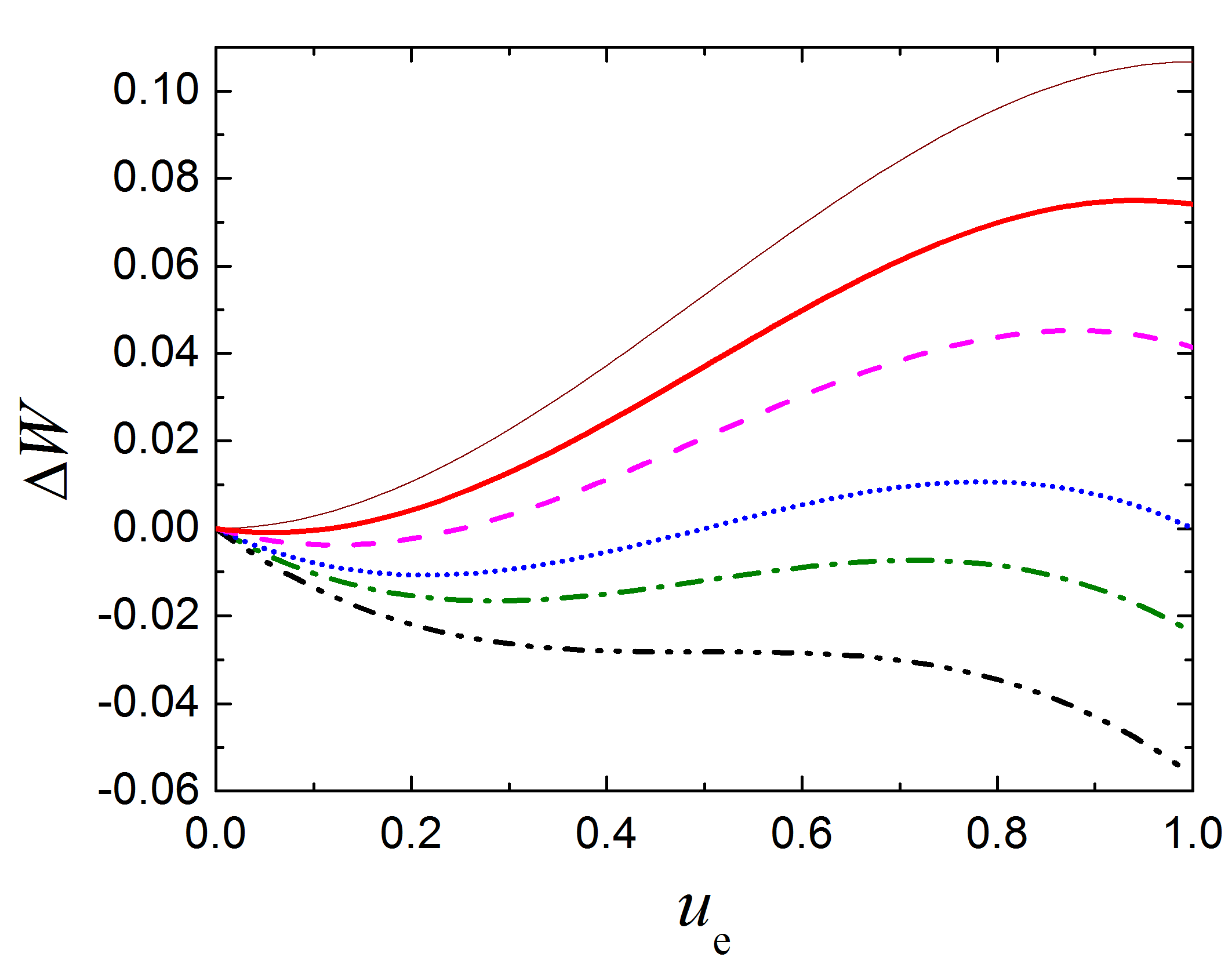}
	\caption{Formation energy of edge stacking dislocations $\Delta W$ (in eV/\AA) in bilayer graphene per unit edge length as a function of relative displacement $u_\mathrm{e}$  (in units of bond length $l$) of atoms of the upper and bottom layers at the edge of the upper layer for angles $\alpha=\phi=\beta = 0^{\circ}$ (Fig.~\ref{fig:statement_edge}). Elongations $\epsilon$ of the bottom layer are 0 (brown thin solid line), $1.10 \cdot 10^{-3}$ (red solid line), $2.21 \cdot 10^{-3}$ (magenta dashed line) , $3.61 \cdot 10^{-3}$ (blue dotted line, critical elongation $\epsilon_\mathrm{c}$),  $4.42 \cdot 10^{-3}$ (green dash-dotted line) and $5.52 \cdot 10^{-3}$ (black dash-dot-dotted line, $\epsilon_\mathrm{max}$).}
	\label{fig:Wedge}
\end{figure}

Using that in bilayer graphene the interlayer interaction energy along the dislocation path, which is the same as the minimum energy path between two adjacent minima AB and AC, can be approximated as \cite{Lebedev2015,Lebedeva2016}
\begin{equation} \label{eq_3}
\begin{split}
	V(u) = V_\mathrm{max}\left(2\cos{\frac{2\pi(u + 1)}{3}} +1\right)^2,
\end{split}
\end{equation} 
we present the formation energy $\Delta W$ of ESDs as a function of the relative displacement $u_\mathrm{e}$ of atoms at the edge as
\begin{equation} \label{eq_4}
\begin{split}
\Delta W(u_\mathrm{e}) =&  -\sqrt{K(\beta)  l^2 V_\mathrm{max}} \Bigg[ u_\mathrm{e} \Bigg(1+ \frac{\epsilon}{\epsilon_\mathrm{max}}  \Bigg) \\
&+ \frac{3}{\pi} \Bigg(\sin{\frac{2\pi(u_\mathrm{e} + 1)}{3}}-\frac{\sqrt{3}}{2} \Bigg) \Bigg],
\end{split}
\end{equation}
where $\epsilon_\mathrm{max}=\sqrt{ K(\beta) V_\mathrm{max}}/k \cos{\alpha} \cos{\phi}$ (we consider the case $\cos{\alpha}, \cos{\phi} >0$). The dependence of the formation energy $\Delta W$ on the relative displacement $u_e$ of atoms at the edge is shown in Fig.~\ref{fig:Wedge}. 
It is seen that at any non-zero elongation of the bottom layer $0 < \epsilon <\epsilon_\mathrm{max}$ this dependence has a minimum at some $0<u_\mathrm{e}=u_\mathrm{o}<1$. This means that an incommensurate structure of the edge with some finite relative displacement of atoms is preferred over the commensurate state, i.e. the ESD is spontaneously formed. The presence of a maximum at $0<u_\mathrm{e}=u_\mathrm{a}<1$, on the other hand, suggest that there is a barrier for transition from the state characterized as the ESD ($u_\mathrm{e}=u_\mathrm{o}<1$) to the state with the complete PSD inside the layers ($u_\mathrm{e}=1$), i.e. penetration of PSDs inside the layers is thermally activated for elongations $0 < \epsilon < \epsilon_\mathrm{max}$. For $\epsilon \ge \epsilon_\mathrm{max}$, the barrier for penetration of PSDs according to the considered model desappears. However, in this case the density of dislocations is no longer low and the interaction of ESDs and PSDs inside the layers cannot be ignored. Therefore, in this limit the considered model is not applicable.

\begin{figure}
	\centering
	\includegraphics[width=\columnwidth]{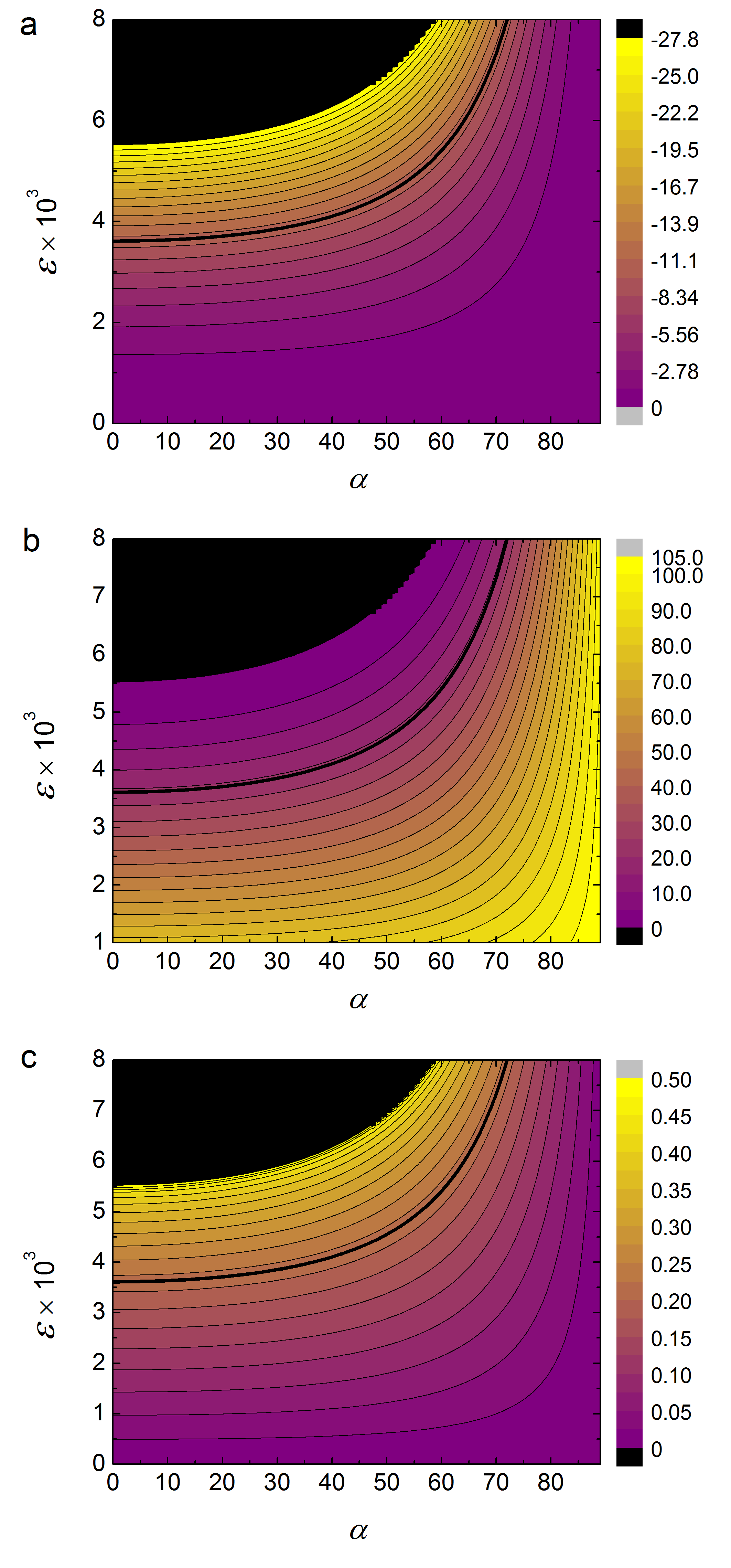}
	\caption{(a) Formation energy $\Delta W_\mathrm{o}$ (in meV/\AA) of edge stacking dislocations in bilayer graphene per unit edge length, (b) activation energy $\Delta W_\mathrm{a}$ (in meV/\AA) for peneration of partial stacking dislocations inside the layers per unit edge length and (c) optimal relative displacement $u_\mathrm{o}$ of atoms of the upper and bottom layers at the edge of the upper layer (in units of the bond length $l$) as functions of the relative elongation  $\epsilon$ of the bottom layer and angle $\alpha$ (in degrees) between the edge Burgers vector and direction elongation (Fig.~\ref{fig:statement_edge}) for the case when the bottom edge is stretched perpendicular to the edge ($\phi=0^{\circ}$ and $\alpha=\beta$). The equipotential lines are shown with a step of (a) -1.39 meV/\AA, (b) 5 meV/\AA~and (c) 0.025. The black area corresponds to elongations exceeding $\epsilon_\mathrm{max}(\alpha)$. The critical elongation  $\epsilon_\mathrm{c}(\alpha)$ is shown by the thick black line.
} 
	\label{fig:energy}
\end{figure}

The displacements $u_\mathrm{o}$ and $u_\mathrm{a}$ follow from the condition   $\partial \Delta W/\partial u_\mathrm{e} = 0$, which is reduced to $\epsilon/\epsilon_\mathrm{max}=V(u_\mathrm{o})/V_\mathrm{max}$,
supplemented by $\partial^2 \Delta W/\partial u_\mathrm{e}^2 > 0$ or $\partial^2 \Delta W/\partial u_\mathrm{e}^2 < 0$, respectively. Using also approximation~(\ref{eq_3}), we find that the optimal relative displacement  $u_\mathrm{o}$ of atoms at the edge of the upper layer can be expressed as (Fig.~\ref{fig:energy}c)
\begin{equation} \label{eq_6}
\begin{split}
 u_\mathrm{o}= \frac{1}{2} - \frac{3}{2\pi}\varphi, \quad \varphi= \arccos{\frac{1}{2}\Bigg(\frac{\epsilon}{\epsilon_\mathrm{max}}  + 1\Bigg)},
\end{split}
\end{equation} 
while the relative displacement $u_\mathrm{a}$ corresponding to the barrier for penetration of PSDs inside the layers is given by $u_\mathrm{a}= 1 - u_\mathrm{o}$. 

The formation energy of ESDs $\Delta W_\mathrm{o}=\Delta W(u_\mathrm{o})$ as a function of the elongation $\epsilon$ of the bottom layer can be found using expression~(\ref{eq_6}) in Eq.~(\ref{eq_4})  (Fig.~\ref{fig:energy}a). The barrier for penetration of PSDs inside the layers is given by the energy difference for the states with $u_\mathrm{e} = u_\mathrm{o}$ and $u_\mathrm{e}=u_\mathrm{a}$ (Fig.~\ref{fig:energy}b)
\begin{equation} \label{eq_8}
\begin{split}
\Delta W_a =  \sqrt{K(\beta) l^2 V_\mathrm{max}} \Bigg[\frac{6}{\pi} \sin{\varphi} - \frac{3\varphi}{\pi} \Bigg(1+ \frac{\epsilon}{\epsilon_\mathrm{max}} \Bigg) \Bigg]
\end{split}
\end{equation} 

Let us now restrict the analysis to the case when the edge is perpendicular to the direction of elongation ($\phi=0^{\circ}$ and $\alpha=\beta$, see Fig.~\ref{fig:struct}). In this case the formation energy $\Delta W_\mathrm{o}$ of ESDs, barrier $\Delta W_\mathrm{a}$ to penetration of PSDs inside the layers and optimal relative displacement $u_\mathrm{o}$ of atoms at the edge are monotonous functions of the elongation $\epsilon$ of the bottom layer and the angle $\alpha$ between the edge Burgers vector and normal to the edge (Fig.~\ref{fig:energy}). The relative displacement $u_\mathrm{o}$ of atoms and the magnitude of the formation energy $\Delta W_\mathrm{o}$ increase with increasing the elongation and decrease upon increasing the the angle $\alpha$ from $0^{\circ}$ to $90^{\circ}$, while the behavior of the barrier $\Delta W_\mathrm{a}$ is opposite. 

Thermodynamic criteria are usually invoked to analyze formation of PSDs inside the layers \cite{Popov2011,Lebedev2015,Lebedeva2016,Bichoutskaia2006, Popov2009}. In particular, the critical elongation $\epsilon_\mathrm{c}$ at which the formation energy of dislocations is zero separates different structural phases in the bilayer, the commensurate phase with no PSDs inside the layers and the structure with a single PSD with a given direction of the Burgers vector \cite{Popov2011,Lebedeva2016}. Using that the critical elongation complies with $\Delta W(u_\mathrm{e}=1) =0$ (see the blue dotted line in Fig.~\ref{fig:Wedge}), we find $\epsilon_\mathrm{c}/\epsilon_\mathrm{max} = 3\sqrt{3}/\pi-1 \approx 0.654$. This ratio does not depend on the angles between the edge Burgers vector, edge and direction of elongation, providing that  the formation energy $\Delta W_\mathrm{o}$ of ESDs, barrier $\Delta W_\mathrm{a}$ to penetration of PSDs inside the layers and optimal relative displacement $u_\mathrm{o}$ of atoms at the edge at the critical elongation are also the same for any angles $\alpha, \beta$ and $\phi$ (Fig.~\ref{fig:energy}). 

Though penetration of PSDs inside the layers becomes energetically favourable at the critical elongation,  the barrier to peneration can still be considerable $\Delta W_\mathrm{a}(\epsilon_\mathrm{c}) = 21$ meV/\AA~ keeping in mind that it should be multiplied by the edge length. In reality, however, it can be expected that penetration of dislocations occurs not along the whole edge at once but gradually. The characteristic length $L$ of the piece of the incommensurate boundary that passes through the saddle-point can be estimated by minimization of the sum $E_\mathrm{a} =E_\mathrm{bar}+ E_\mathrm{el}$ of the energy $E_\mathrm{bar}=\Delta W_\mathrm{a}L$ required to pass the barrier and elastic energy $E_\mathrm{el}$ associated with the shear deformation needed to bend the boundary. The latter quantity can be estimated from the density of the elastic energy $Gl^2 (u_\mathrm{a}-u_\mathrm{o})^2/4L^2$ multiplied by the area involved $L l_\mathrm{D} u_\mathrm{a}$, where $l_\mathrm{D} = \sqrt{kl^2/4V_\mathrm{max}}$ is the characteristic width of the boundary between commensurate domains \cite{Popov2011,Lebedev2015,Lebedeva2016}. In this way we estimate
\begin{equation} \label{eq_9}
\begin{split}
L \sim \sqrt{\frac{kl^2u_\mathrm{a}l_\mathrm{D}}{8(1+\nu)\Delta W_\mathrm{a}}}(u_\mathrm{a}-u_\mathrm{o})
\end{split}
\end{equation} 
 and get $L \sim 90$~\AA, which corresponds to the total activation energy $E_\mathrm{a}  \sim 4$~eV.

It should be also mentioned that so far we have considered the angle $\alpha$ between the edge Burgers vector and direction of elongation as a parameter. In a given graphene sample only six descrete directions of the edge Burgers vector corresponding to the armchair directions are possible \cite{Lebedeva2016}.  Furthermore, due to the hexagonal symmetry of the potential energy surface there are only three ways out of each potential energy minimum. This corresponds in general to two possible directions of the edge Burgers vector for one edge and one direction for the other. However, as follows from the analysis above (Fig.~\ref{fig:energy}a,b), in the case then the edges are perpendicular to the direction of elongation, the ESDs with the smallest possible angle $\alpha$ are preferred and their formation and penetration of the corresponding PSDs inside the layers should be observed. 
  
\section{Conclusions and discussion}
A new phenomenon,  spontaneous formation of edge stacking dislocations, is proposed to take place in two-dimensional hexagonal bilayers with the shorter upper layer upon stretching the bottom layer. The formation energy of these defects and the deviation from the commensurate structure are found to become more and more pronounced upon increasing the external strain. Recent advances in experimental characterisation of partial stacking dislocations in bilayer and few-layer graphene by scanning transmission microscopy \cite{Alden2013,Lin2013} and scanning tunneling microscopy \cite{Yankowitz2014} give us hope that these new structures could be soon observed experimentally.

Our estimates also show that there is a considerable barrier for penetration of dislocations inside the layers, which can be reduced by increasing the elongation of the bottom layer. Taking this barrier into account is important for observation of the commensurate-incommensurate pha-se transitions in two-dimensional atomically thin bilayers or double-walled nanotubes.

Measurements of the characteristic width of partial stacking dislocations spontaneously formed in bilayer gra-phene have been used to estimate the barrier to relative motion of graphene layers \cite{Alden2013}. The value obtained of 2.4 meV/atom (per atom of the adsorbed layer), however, is different from the results of most of the density functional theory calculations (see \cite{Lebedeva2016a} and references therein) and barrier of 1.7 meV/atom deduced from the experimental data on the shear mode frequency in bilayer and few-layer graphene and graphite \cite{Popov2012}. There are two complications that can limit the accuracy of barrier estimates from observations of partial stacking dislocations. First, the partial stacking dislocation is a soliton and the dislocation width should be determined through the analysis of derivative $\mathrm{d} u/\mathrm{d} x$ of the relative displacement $u$ of the layers in the direction $x$ perpendicular to the boundary between commensurate domains \cite{Popov2011, Lebedev2015, Bichoutskaia2006, Popov2009}. However, at the moment precise measurements of this derivative are not possible. Second, the presence of both tensile and shear strains in the system leads to different angles between the Burger vector and the boundary between commensurate domains so that the dependence of the dislocation width on this angle \cite{Alden2013, Lebedev2015} should be taken into account. 

Based on the results obtained in the present paper we can suggest another experimental scheme for estimation of the barrier to relative motion of graphene layers that does not suffer from these difficulties. Namely we propose to place an armchair graphene ribbon on a graphene layer in the commensurate relative orientation and to apply a controlled relative elongation to the layer along the ribbon leaving the ribbon free. In this case the purely tensile edge stacking dislocation is formed at one of the ribbon edges for any nonzero relative elongation of the stretched layer and the barrier can be found by measurements of the displacement of the ribbon edge at different elongations. It should be noted that changes of electronic properties of monolayer graphene deposited on the flexible substrate have been studied for uniaxial strains in the subtrate up to 0.8\% \cite{Ni2008}. A similar approach can be used to stretch the graphene layer with the graphene ribbon adsorbed  and to obtain displacements of edge atoms at the ribbon ends as functions of the strain in the stretched layer. 

The observation of edge stacking dislocations can be also used to analyze small strains generated in graphene layers upon adsorption on a substrate (see, for example, Ref. \cite{Dahal2014} for a review) or covalent functionalization \cite{Milowska2013}. Since there is a lattice mismatch in the adsorbed or chemically modified and free graphene layers, formation of edge stacking dislocations should take place at edges of the free layer upon placing it on the adsorbed or chemically modified layer. For a lattice mismatch of the layers smaller than the critical one, displacements of atoms at edges of the free layer are unambiguously determined by the mismatch. Thus measurements of these displacements can be used to study changes in lattice constants of adsorbed or chemically modified graphene layers. Such an approach can be particularly useful in the cases of strained substrate \cite{Weng2010} and multilayer graphene \cite{Zhu2015} with the thickness dependent strain or for determination of the average lattice constant of the graphene membrane chemically modified at one side.

\section*{Acknowledgments}
The authors acknowledge the Russian Foundation of Basic Research (14-02-00739-a). IL acknowledges the financial support from Grupos Consolidados UPV/EHU del Gobierno Vasco (IT578-13) and H2020-NMP-2014 project ``MOSTOPHOS" (n. 646259).

\section*{References}
\bibliography{rsc}

\end{document}